\documentclass[11pt,showpacs]{revtex4}
\input epsf.sty

\usepackage{ulem}
\usepackage{cancel}
\usepackage{color}

\def\red#1{\textcolor{red}{#1}}

\def\comment#1{}

\def\E{{\mathcal E}}

\usepackage{graphicx}
\begin{document}
\title{Vectorlike $W^\pm$-boson coupling at TeV and third family fermion masses 
\\ 
}
\author{She-Sheng Xue}
\email{xue@icra.it}
\affiliation{ICRANeT, Piazzale della Repubblica, 10-65122, Pescara,\\
Physics Department, University of Rome ``La Sapienza'', Rome,
Italy} 


\begin{abstract}
In the third fermion family 
and gauge symmetry of the Standard Model (SM), 
we study the quark-quark, lepton-lepton and quark-lepton 
four-fermion operators in an effective 
theory at high energies. These operators have nontrivial 
contributions to the Schwinger-Dyson equations for fermion self-energy 
functions and the $W^\pm$-boson coupling vertex. 
As a result, the top-quark mass is generated via the spontaneous 
symmetry breaking of $\langle\bar t t \rangle$-condensate and 
the $W^\pm$-boson coupling becomes approximately vectorlike  
at TeV scale. The bottom-quark, tau-lepton and tau-neutrino 
masses are generated via the explicit symmetry breaking 
of $W^\pm$-contributions and quark-lepton interactions.
Their masses and Yukawa couplings are functions of the top-quark mass and 
Yukawa coupling. 
We qualitatively show the hierarchy of fermion masses and Yukawa couplings 
of the third fermion family. We also discuss the possible collider signatures 
due to the vectorlike (parity-restoration) feature of $W^\pm$-boson coupling at high energies.
\comment{We also give some speculative discussions on the resonance 
channels of massive (TeV) composite fermions. They can decay to
$WW$, $WZ$ and $ZZ$ two boson-tagged jets with additional quark jets.  
If the recent ATLAS and CMS results of resonances in diboson channel with invariant
masses around 2.0 TeV are confirmed, these resonances are also
expected in the channels of four quark jets, 
the latter should have larger rate. Moreover, at the same 
energy TeV scale, the resonances of massive composite fermions are 
formed by high-energy sterile neutrino (dark matter) collisions and decay, 
leading to an excess of cosmic ray particles in space, producing 
signals of SM particles in underground.}  
\end{abstract}


\maketitle

\section
{\bf Introduction}
\hskip0.1cm
The parity-violating (chiral) gauge symmetries and spontaneous/explicit 
breaking of these symmetries for the hierarchy of fermion masses and
mixing angles have 
been at the center of a conceptual elaboration that has played a major 
role in donating to mankind the beauty of the SM for fundamental
particle physics. On the one hand the {\it composite} 
Higgs-boson model or the Nambu-Jona-Lasinio (NJL) \cite{njl} with
four-fermion operators, and on the other  
the phenomenological model \cite{higgs}
of the {\it elementary} Higgs boson, 
they are effectively equivalent for the SM at low energies and provide an elegant 
and simple description for the electroweak symmetry 
breaking and intermediate gauge boson masses. 
The experimental results of Higgs-boson mass 126 GeV \cite{ATLASCMS} and top-quark mass 173 GeV \cite{CDFD0} begin to shed light 
on this most elusive and fascinating arena of fundamental particle physics. 

In order to accommodate high-dimensional operators of fermion 
fields in the SM framework of a well-defined quantum field theory at the 
high-energy scale $\Lambda$, it is essential and necessary to study: 
(i) what physics beyond the SM at the scale $\Lambda$ explains the 
origin of these operators; (ii) which dynamics of these operators 
undergo in terms of their dimensional couplings (e.g., $G$) 
and energy 
scale $\mu$; (iii) associating to these dynamics, where infrared (IR) 
and ultraviolet (UV) 
stable fixed points of these couplings locate and what characteristic energy 
scale is; (iv) in the IR domain and 
UV domain (scaling regions) of these stable IR and UV fixed points, 
which operators become physically relevant (effectively dimension-4) 
and renormalizable following renormalization group (RG) equations (scaling laws), 
and other irrelevant operators are suppressed by the cutoff at least 
$\mathcal O(\Lambda^{-2})$.    

The strong technicolor dynamics of extended gauge theories at the 
TeV scale was invoked \cite{hill1994,bhl1990a}
to have a natural scheme incorporating the 
relevant four-fermion operator 
\begin{eqnarray} 
G(\bar\psi^{ia}_Lt_{Ra})(\bar t^b_{R}\psi_{Lib})
\label{bhl}
\end{eqnarray}
of the $\langle\bar t t\rangle$-condensate model of Bardeen, Hill and Lindner (BHL) \cite{bhl1990} and 
notations will be given later in Eq.~(\ref{bhlx}). 
This relevant four-fermion operator (\ref{bhl}) undergoes 
the dynamics of spontaneous symmetry breaking (SSB) in 
the IR domain (small $G\gtrsim G_c$) of infrared
fixed point $G_c$ (critical value) associated with the SSB and 
characteristic energy scale (vev) $v\approx 239.5$ GeV. 
The analysis of this composite Higgs boson model was made \cite{bhl1990} to show 
the low-energy effective Lagrangian, RG equations,
the composite Goldstone modes (pseudoscalars $\bar\psi\gamma_5\psi$) 
for the longitudinal modes of massive $W^\pm$ and $Z^0$ gauge bosons, 
and the composite scalar ($H\sim\bar\psi\psi$) for the Higgs boson in the SM. 
On the other hand, these relevant operators can be constructed 
on the basis of phenomenology of the SM at low-energies.  
In 1989, several authors \cite{bhl1990,nambu1989,Marciano1989} 
suggested that the symmetry breakdown of the SM could be 
a dynamical mechanism of the NJL type that 
intimately involves the top quark at the high-energy scale $\Lambda$. 
Since then, many models based on this idea have been 
studied \cite{DSB_review}. The low-energy SM physics
was supposed to be achieved by the RG equations in the IR domain 
of the IR-stable fixed point 
with $v\approx 239.5$ GeV \cite{bhl1990a,Marciano1989,bhl1990}. 

Nowadays, the top-quark and Higgs boson masses are known and 
they completely determine the boundary 
conditions for the RG equations of the composite Higgs boson model \cite{bhl1990}.
Using the experimental values of top-quark and Higgs boson masses, 
we obtained \cite{xue2013,xue2014} the unique solutions to these RG equations 
provided the appropriate nonvanishing 
form factor of the composite Higgs boson in TeV scales where the effective 
quartic coupling of composite Higgs bosons vanishes.

The form factor of composite Higgs boson $H\sim(\bar\psi\psi)$ is finite and  
does not vanish in the SSB phase (composite Higgs phase for small $G\gtrsim G_c$), 
indicating that the tightly bound composite Higges particle behaves 
as if an elementary particle. On the other hand, 
due to large four-fermion coupling $G$, massive composite fermions $\Psi \sim (H\psi)$ 
are formed by combining a composite Higgs boson $H$ with an elementary fermion $\psi$ 
in the symmetric phase where the SM gauge symmetries are exactly preserved \cite{xue1997}. 
This indicates that a second-order phase transition from the SSB phase to the 
SM gauge symmetric phase takes place at the critical point $G_{\rm crit}> G_c$. 
In addition the effective quartic coupling of composite Higgs bosons vanishing 
at $\E\sim$ TeV scales indicates the characteristic energy scale of such phase transition. 
The energy scale $\E$ is much lower than the cutoff scale 
$\Lambda$ ($\E \ll \Lambda$) so that the fine-tuning (hierarchy) problem 
of fermion masses $m_f\ll \Lambda$ or the pseudoscalar 
decay constant $f_\pi\ll \Lambda$ can be  avoided \cite{xue2013}. 

\comment{In Ref.~\cite{xue2014}, we preliminarily calculated the $\beta(G)$-function and showed 
$\beta(G)>0$ for small coupling $G$ and $\beta(G)<0$ for large coupling $G$, indicating
the critical point $G_{\rm crit}$ to be a UV-stable fixed point. 
In the UV domain (large $G\gtrsim G_{\rm crit}$) 
of UV-stable fixed point $G_{\rm crit}$, an effective 
field theory of composite particles can be defined with a characteristic energy scale $\E\sim$ TeV scales, and four-fermion operators (\ref{bhl}) 
are expected to acquire anomalous dimensions and thus become relevant (renormalizable)
operators of effective dimension-4. 
The effective field theory preserves the SM gauge symmetries, however 
its composite particle spectrum and parity-conservation are 
completely different from the SM. This is reminiscent of the 
asymptotic safety \cite{w1} that quantum 
field theories regularized at UV cutoff $\Lambda$ 
might have a non-trivial 
UV-stable fixed point, RG flows 
are attracted into the UV-stable fixed point  
with a finite number of physically renormalizable operators. 
The small or large 
four-fermion coupling $G$ brings us into the IR domain or the UV domain, they
are two distinct domains. 
This lets us recall the QCD dynamics: asymptotically 
free quark states in the UV domain of an UV-stable fixed point 
and bound hadron states in the IR domain of a possible IR-stable fixed point.
}
In this article, after a short review that recalls and 
explains the quantum-gravity origin 
of four-fermion operators at the cutoff $\Lambda$, 
the SSB and $\langle \bar t t\rangle$-condensate model, 
we show that due to four-fermion operators
(i) there are the SM gauge symmetric vertexes 
of quark-lepton interactions; 
(ii) the one-particle-irreducible (1PI) 
vertex function of $W^\pm$-boson coupling becomes 
approximately vectorlike at TeV scale. 
Both interacting vertexes contribute the explicit symmetry breaking (ESB) 
terms to Schwinger-Dyson (SD) equations for fermion self-energy functions. 
As a result, once the top-quark mass is generated via the SSB, 
other fermion $(\nu_\tau, \tau, b)$ masses are generated by the ESB via 
quark-lepton interactions and $W^\pm$-boson vectorlike coupling. 
In the third fermion family, we qualitatively show the hierarchy of fermion masses 
and effective Yukawa couplings 
in terms of the top-quark mass 
and Yukawa coupling. In the concluding section, a summary of basic points of the scenario and 
its extension to three fermion families is given. In addition, we present some 
discussions on the possible experimental relevance of running Yukawa couplings obtained and 
parity-conservation feature of the $W^\pm$-boson coupling at TeV scale \footnote{More discussions on the experimental aspects of this scenario can be found in the Refs.~S.-S.~Xue,
arXiv1601.06845 and arXiv1506.05994v2. The latter contains the part of this article.}.    

\comment{
On the other hand, recall the discussions on 
relevant four-fermion operators in the both domains of IR- and UV-fixed points, 
as well as their 
resonant and nonresonant new phenomena for experimental searches \cite{xue2015}.
In the last section, we
discuss the experimental relevance of effective Yukawa and 
vectorlike $W^\pm$-couplings.
In the view of the recent ATLAS and CMS results \cite{ATLAS2015,CMS2015}, 
we also discuss the channels of massive (TeV) composite Dirac fermions 
decaying to: 
(i) the event of two boson-tagged jets with additional quark jets, 
(ii) the event of four quark jets, and their possible 
correlation and experimental consequences. To end this article, we point out 
the scattering 
channels of high-energy sterile neutrinos (dark matter) to produce the resonances 
of massive composite fermions that decay to SM particles at the TeV scale in the 
space and underground. 
}

\noindent
\section {Four-fermion operators from quantum gravity}
\hskip0.1cm
A well-defined quantum field theory for the SM Lagrangian requires a natural 
regularization (cutoff $\Lambda$) fully preserving the SM chiral-gauge 
symmetry. The quantum gravity naturally provides a such regularization of discrete 
space-time with the minimal length $\tilde a\approx 1.2\,a_{\rm pl}$ \cite{xue2010}, 
where the Planck length 
$a_{\rm pl}\sim 10^{-33}\,$cm and scale $\Lambda_{\rm pl}=\pi/a_{\rm pl}\sim 10^{19}\,$GeV. 
However, the no-go theorem \cite{nn1981} tells us 
that there is not any consistent way to regularize 
the SM bilinear fermion Lagrangian to exactly preserve the SM chiral-gauge symmetries, which 
must be explicitly broken at the scale of fundamental space-time cutoff $\tilde a$. 
This implies that the natural quantum-gravity regularization for the SM should
lead us to consider at least dimension-6 four-fermion operators originated from quantum-gravity 
effects at short distances \footnote{In the regularized and quantized EC theory \cite{xue2010}
with a basic space-time cutoff, in addition to dimension-6 four-fermion operators,
there are high-dimensional fermion operators ($d>6$), 
e.g., $\partial_\sigma J^\mu\partial^\sigma J{_\mu}$, which are suppressed
at least by ${\mathcal O}(\tilde a^4)$.
}. 

On the other hand, it is known that four-fermion operators of the classical and 
torsion-free Einstein-Cartan (EC) theory are naturally
obtained by integrating over ``static'' torsion fields at the Planck length,  
\begin{eqnarray}
{\mathcal L}_{EC}(e,\omega,\psi)&=&  {\mathcal L}_{EC}(e,\omega) + 
\bar\psi e^\mu {\mathcal D}_\mu\psi +GJ^dJ_d,
\label{ec0}
\end{eqnarray}
where the gravitational Lagrangian ${\mathcal L}_{EC}={\mathcal L}_{EC}(e,\omega)$, 
tetrad field $e_\mu (x)= e_\mu^{\,\,\,a}(x)\gamma_a$,
spin-connection field $\omega_\mu(x) = \omega^{ab}_\mu(x)\sigma_{ab}$,  
the covariant derivative ${\mathcal D}_\mu =\partial_\mu - ig\omega_\mu$ and 
the axial current $J^d=\bar\psi\gamma^d\gamma^5\psi$ of massless fermion fields. 
The four-fermion coupling $G$ relates to the gravitation-fermion gauge 
coupling $g$ and fundamental space-time cutoff $\tilde a$. 

Within the SM fermion content, we consider massless left- and 
right-handed Weyl fermions $\psi^f_{_L}$ and $\psi^f_{_R}$ carrying 
quantum numbers of the SM symmetries, as well as three
right-handed Weyl sterile neutrinos 
$\nu^f_{_R}$ and their left-handed conjugated fields 
$\nu^{f\, c}_{_R}=i\gamma_2(\nu_{_R})^*$, where ``$f$'' is the fermion-family index.  
Analogously to the EC theory (\ref{ec0}), we obtain a torsion-free, 
diffeomorphism and {\it local} gauge-invariant 
Lagrangian 
\begin{eqnarray}
{\mathcal L}
&=&{\mathcal L}_{EC}(e,\omega)+\sum_f\bar\psi^f_{_{L,R}} e^\mu {\mathcal D}_\mu\psi^f_{_{L,R}} 
+ \sum_f\bar\nu^{f c}_{_{R}} e^\mu {\mathcal D}_\mu\nu^{f c}_{_{R}}\nonumber\\
&+&G\left(J^{\mu}_{_{L}}J_{_{L,\mu}} + J^{\mu}_{_{R}}J_{_{R,\mu}} 
+ 2 J^{\mu}_{_{L}}J_{_{R,\mu}}\right)\nonumber\\
&+&G\left(j^{\mu}_{_{L}}j_{_{L,\mu}} + 2J^{\mu}_{_L}j_{_{L,\mu}} 
+ 2 J^{\mu}_{_R}j_{_{L,\mu}}\right),
\label{art}
\end{eqnarray}
where we omit the gauge interactions in ${\mathcal D}_\mu$ 
and axial currents read
\begin{eqnarray}
J^{\mu}_{_{L,R}}\equiv \sum_f\bar\psi^f_{_{L,R}}\gamma^\mu\gamma^5\psi^f_{_{L,R}}, \quad
j^{\mu}_{_L}\equiv \sum_f\bar\nu^{fc}_{_R}\gamma^\mu\gamma^5\nu^{fc}_{_R}. 
\label{acur}
\end{eqnarray}
The four-fermion coupling $G$ is unique for all four-fermion operators and 
high-dimensional fermion operators ($d>6$) are neglected. 
If torsion fields that couple to fermion fields 
are not exactly static, propagating a short distance 
$\tilde \ell \gtrsim \tilde a$, 
characterized by their large masses 
$\Lambda\propto \tilde \ell^{-1}$, this implies the four-fermion 
coupling $G\propto \Lambda^{-2}$.
We will in the future 
address the issue of how the space-time cutoff $\tilde a$ due to quantum 
gravity relates to the cutoff scale $\Lambda(\tilde a)$ 
possibly by intermediate torsion fields 
or the Wilson-Kadanoff renormalization group approach.
In this article, we adopt the effective four-fermion operators (\ref{art}) 
in the context of a well-defined quantum field theory 
at the high-energy scale $\Lambda$.

By using the Fierz theorem \cite{itzykson}, the dimension-6 four-fermion operators 
in Eq.~(\ref{art}) can be written as  
\begin{eqnarray}
&+&(G/2)\left(J^{\mu}_{_{L}}J_{_{L,\mu}} + J^{\mu}_{_{R}}J_{_{R,\mu}} 
+ j^{\mu}_{_{L}}j_{_{L,\mu}} + 2J^{\mu}_{_L}j_{_{L,\mu}}\right)\label{art0}\\
&-&G\sum_{ff'}\left(\, \bar\psi^f_{_L}\psi^{f'}_{_R}\bar\psi^{f'}_{_R} \psi^f_{_L}
+\, \bar\nu^{fc}_{_R}\psi^{f'}_{_R}\bar\psi^{f'}_{_R} \nu^{fc}_{_R}\right),
\label{art0'}
\end{eqnarray}
which preserve the SM gauge symmetries.
Equations (\ref{art0}) and 
(\ref{art0'}) represent repulsive and attractive operators respectively. 
In Ref.~\cite{xue2015}, we pointed out that the repulsive 
four-fermion operators (\ref{art0}) 
are suppressed by the cutoff ${\mathcal O}(\Lambda^{-2})$, and cannot become 
relevant and renormalizable operators of effective dimension-4 
in the IR domain where the SSB dynamics occurs. 
\comment{
Instead, four-fermion operators (\ref{art0'}) are
relevant operators in the both domains of IR- and UV-stable fixed points, respectively 
associated with the SSB dynamics and formation of composite fermions. 
We presented some discussions of these relevant operators and their 
resonant and nonresonant new phenomena for experimental searches. 
We will proceed further discussions in the last section 
of this article. 
}

Thus the torsion-free EC theory with the relevant 
four-fermion operators read,
\begin{eqnarray}
{\mathcal L}
&=&{\mathcal L}_{EC}+\sum_{f}\bar\psi^f_{_{L,R}} e^\mu {\mathcal D}_\mu\psi^f_{_{L,R}} 
+ \sum_{f}\bar\nu^{ fc}_{_{R}} e^\mu {\mathcal D}_\mu\nu^{ fc}_{_{R}}\nonumber\\
&-&G\sum_{ff'}\left(\, \bar\psi^{f}_{_L}\psi^{f'}_{_R}\bar\psi^{f'}_{_R} \psi^{f}_{_L}
+\, \bar\nu^{fc}_{_R}\psi^{f'}_{_R}\bar\psi^{f'}_{_R} \nu^{fc}_{_R}\right)+{\rm h.c.},
\label{art1}
\end{eqnarray}
where the two-component Weyl fermions $\psi^{f}_{_L}$ and $\psi^{f}_{_R}$  
respectively are the $SU_L(2)\times U_Y(1)$ gauged doublets and singlets of the SM. 
For the sake of compact notations, $\psi^{f}_{_R}$ is also used to represent 
$\nu^f_R$, which has no SM quantum numbers. 
All fermions are massless, 
they are four-component Dirac fermions 
$\psi^f=(\psi_L^f+\psi_R^f)$, two-component right-handed Weyl neutrinos $\nu^f_L$ 
and four-component sterile Majorana neutrinos $\nu_M^f=(\nu_R^{fc}+\nu_R^f)$ whose 
kinetic terms read
\begin{eqnarray}
\bar\nu^f_{_{L}} e^\mu {\mathcal D}_\mu\nu^f_{_{L}}, \quad
\bar\nu_{_{M}}^f e^\mu {\mathcal D}_\mu\nu_{_{M}}^f =
\bar\nu^f_{_{R}} e^\mu {\mathcal D}_\mu\nu^f_{_{R}} 
+ \bar\nu^{ fc}_{_{R}} e^\mu {\mathcal D}_\mu\nu^{ fc}_{_{R}}.
\label{mnu}
\end{eqnarray} 
In Eq.~(\ref{art1}), $f$ and $f'$ ($f,f'=1,2,3$) are 
fermion-family indexes summed over respectively for three 
lepton families (charge $q=0,-1$) and three quark families ($q=2/3,-1/3$). 
Equation (\ref{art1}) preserves not only the SM gauge symmetries 
and global fermion-family symmetries, but also the global symmetries for 
fermion-number conservations. 
\comment{Relating to the gravitation-fermion gauge 
coupling $g$, the effective four-fermion coupling $G$ is unique 
for all four-fermion operators, and its strength depends on energy scale and 
characterizes: (i) the domain of IR fixed point where the 
spontaneous breaking of SM gauge-symmetries occurs (see for example 
\cite{bhl1990}) 
and (ii) the domain of UV fixed point where the SM gauge-symmetries 
are restored and massive (TeV) composite Dirac fermions are formed \cite{xue2014}. 
}

\section
{\bf The third fermion family}
\hskip0.1cm
In this section, 
we discuss how the quark and lepton Dirac mass matrices are generated by the 
SSB via four-fermion operators. 
In Eq.~(\ref{art1}), the four-fermion operators of the quark sector are  
\begin{eqnarray}
-G\, \sum_{ff'}\bar\psi^{f}_{_L}\psi^{f'}_{_R}\bar\psi^{f'}_{_R} \psi^{f}_{_L}.
\label{q}
\end{eqnarray}
Due to the unique four-fermion coupling $G$ 
and the global fermion-family $U_L(3)\times U_R(3)$ symmetry 
of Eq.~(\ref{q}), 
we perform chiral transformations ${\mathcal U}_L\in U_L(3)$ and 
${\mathcal U}_R\in U_R(3)$ so that $f=f'=1,2,3$, the four-fermion operator (\ref{q})
is only for each quark family 
and all quark fields are Dirac mass eigenstates. 
The four-fermion operators (\ref{q}) read, 
\begin{eqnarray}
G\left[(\bar\psi^{ia}_Lt_{Ra})(\bar t^b_{R}\psi_{Lib})
+ (\bar\psi^{ia}_Lb_{Ra})(\bar b^b_{R}\psi_{Lib})\right]+{\rm ``terms"},
\label{bhlx}
\end{eqnarray}
where $a,b$ and $i,j$ are the color and flavor indexes 
of the top and bottom quarks, the left-handed quark 
doublet 
$\psi^{ia}_L=(t^{a}_L,b^{a}_L)$ 
and the right-handed singlet $\psi^{a}_R=t^{a}_R,b^{a}_R$.
The first and second terms in Eq.~(\ref{bhlx}) are respectively 
the four-fermion operators of top-quark channel \cite{bhl1990} 
and bottom-quark channel,
whereas ``terms" stands for 
the first and second quark families that can be obtained 
by substituting $t\rightarrow u,c$ and $b\rightarrow d,s$. 

In Eq.~(\ref{art1}), the four-fermion operators relating to the 
lepton Dirac mass matrix are  
\begin{eqnarray}
-G\, \sum_{ff'}\left[\bar\ell^{f}_{_L}\ell^{f'}_{_R}\bar\ell^{f'}_{_R} \ell^{f}_{_L}+
(\bar\ell^{f}_L\nu^{f'}_{R})(\bar \nu^{f'}_{R}\ell^f_{L})\right],
\label{l}
\end{eqnarray}
where Dirac lepton fields $\ell^{f}_{_L}$ and 
$\ell^{f}_{_R}$ are the SM $SU_L(2)$ doublets and singlets
respectively, and $\nu^{f}_{R}$ are three sterile neutrinos. 
Analogously to the quark sector (\ref{q}), 
we perform chiral transformations ${\mathcal U}_L\in U_L(3)$ and 
${\mathcal U}_R\in U_R(3)$ so that $f=f'$, the four-fermion operators (\ref{l})
are only for each lepton family 
and all lepton fields are Dirac mass eigenstates. Namely, 
the four-fermion operators (\ref{l}) become
\begin{eqnarray}
G\sum_{\ell}\left[(\bar\ell^{i}_L\ell_{R})(\bar \ell_{R}\ell_{Li})
+ (\bar\ell^{i}_L\nu^\ell_{R})(\bar \nu^\ell_{R}\ell_{Li}) 
\right],
\label{bhlxl}
\end{eqnarray}
where three right-handed sterile neutrinos $\nu^\ell_R$ 
($\ell=e,\mu,\tau$), the left-handed lepton 
doublets $\ell^i_L=(\nu^\ell_L,\ell_L)$ and the right-handed singlets 
$\ell_{R}$.
\comment{ and the conjugate fields of sterile neutrinos 
$\nu_R^{\ell c}=i\gamma_2(\nu_R^{\ell})^*$.  
Coming from the second term in Eq.~(\ref{art1}), the last term in Eq.~(\ref{bhlxl}) 
preserves the symmetry 
$U_{\rm lepton}(1)$ for the lepton-number 
conservation, although $(\bar \nu^{\ell}_{R}\nu^{\ell c}_{R})$ 
violates the lepton number of family ``$\ell$'' by two units. 
}

In the IR domain of the SM, the four-fermion coupling $G\gtrsim G_c$ and the SSB leads to 
the fermion-condensation $M_{ff'} =-G\langle \bar \psi^f \psi^{f'}\rangle=m\delta_{ff'}\not=0$, 
two diagonal mass matrices of quark sectors $q=2/3$ and $q=-1/3$
satisfying $3+3$ mass-gap equations. 
It was demonstrated \cite{xue2013_1} that as an energetically favorable 
solution of the SSB ground state of the SM, 
only top-quark is massive ($m^{\rm sb}_t=- G\langle \bar\psi_t \psi_t\rangle\not=0$), 
otherwise there would be more Goldstone modes 
in addition to those becoming the longitudinal modes of massive gauge bosons.
In other words, among four-fermion operators (\ref{bhlx}) 
and (\ref{bhlxl}), the $\langle\bar t t\rangle$-condensate model (\ref{bhl})
is the unique channel undergoing the SSB of SM gauge symmetries, 
for the reason that this is energetically favorable, i.e., 
the ground-state energy is minimal when   
the maximal number of Goldstone modes are three and equal to the number of
the longitudinal modes of massive gauge bosons in the SM.  
Moreover,
the four-fermion operators (\ref{l}) of the lepton sector do not  
undergo the SSB leading to the lepton-condensation 
$M_{ff'} =-G\langle \bar \ell^f \ell^{f'}\rangle=m_\ell\delta_{ff'}\not=0$, i.e., 
two diagonal mass matrices of the lepton sector ($q=0$ and $q=-1$).
The reason is that the effective four-lepton coupling $(GN_c)/N_c$ is $N_c$-times smaller 
than the four-quark coupling $(GN_c)$, where the color number $N_c=3$.
In the IR domain $(G\gtrsim G_c)$ of the IR-stable fixed point $G_c$ 
(the critical value), the effective four-quark coupling is above the critical value and 
the SSB occurs, whereas the effective four-lepton coupling is below 
the critical value and the SSB does not occur. 

As a result, only the top quark acquires its mass via the SSB and
four-fermion operator (\ref{bhl}) of the top-quark channel 
becomes the relevant operator following the RG equations 
in the IR domain \cite{bhl1990}. 
While all other quarks and leptons do not acquire their masses via the SSB 
and their four-fermion operators 
(\ref{q}) and (\ref{l}) are irrelevant dimension-6 operators, 
whose tree-level amplitudes of four-fermion scatterings are 
suppressed ${\mathcal O}(\Lambda^{-2})$, thus their deviations 
from the SM are experimentally inaccessible \cite{xue2015}. However they 
acquire their masses because their SD equations acquire 
the ESB induced by the $W^\pm$-boson vectorlike coupling 
and quark-lepton interactions, see Secs.~\ref{ESBS} and \ref{SDS}.  
It is difficult to analyze the SD equations of three fermion families
all together. Beside, the fermion masses in the third fermion family 
are much heavier than those in the first or second fermion family, 
and the off-diagonal element is much smaller than the diagonal one 
in the family mixing matrices, like the CKM one. 
For these reasons and observations, to the leading order of 
approximation, we focus on the third fermion family in this article so as to
first qualitatively explain and show how the bottom quark, tau lepton and tau 
neutrino acquire their masses as functions of the top-quark mass.

\begin{figure}
\begin{center}
\includegraphics[height=1.40in]{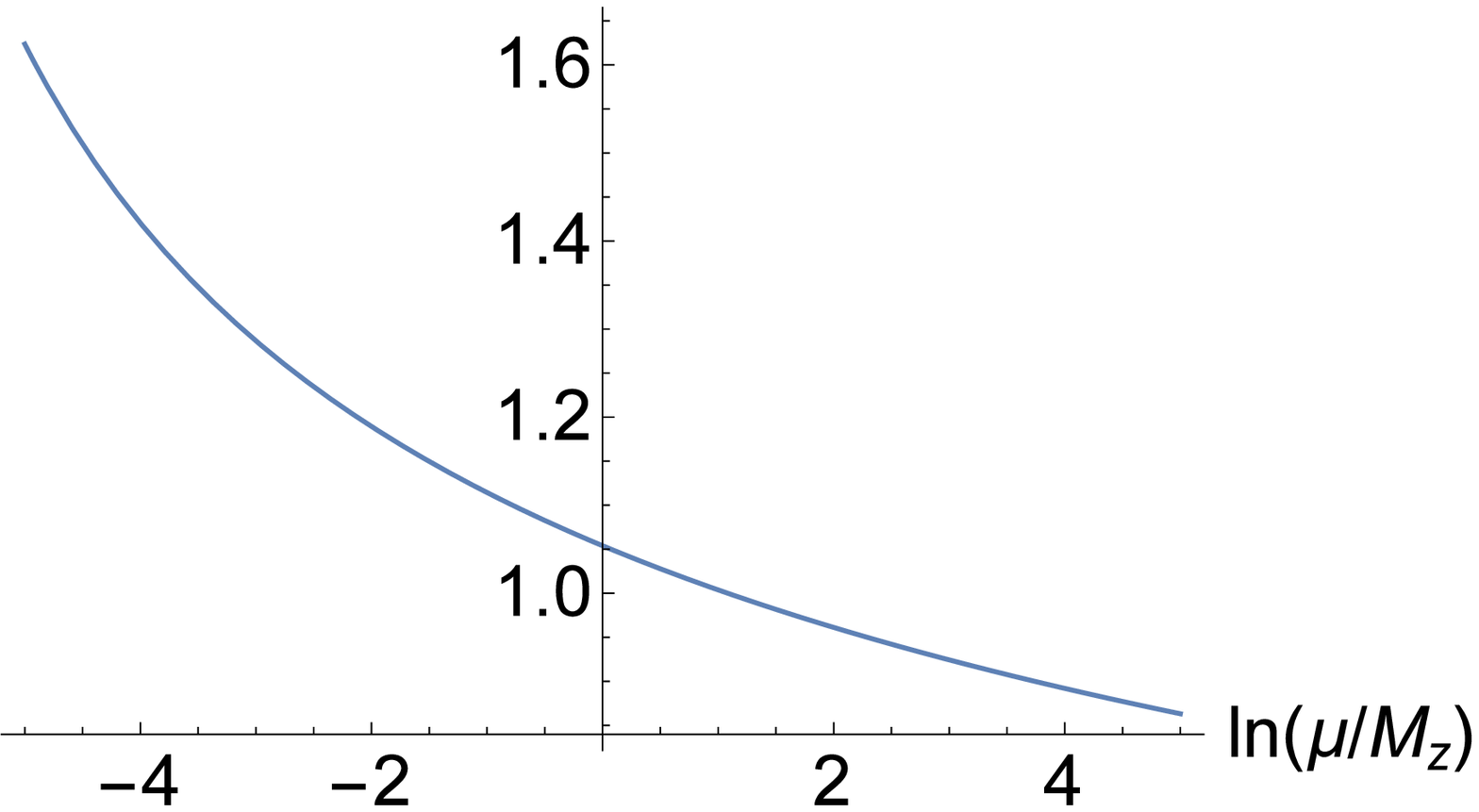}
\includegraphics[height=1.40in]{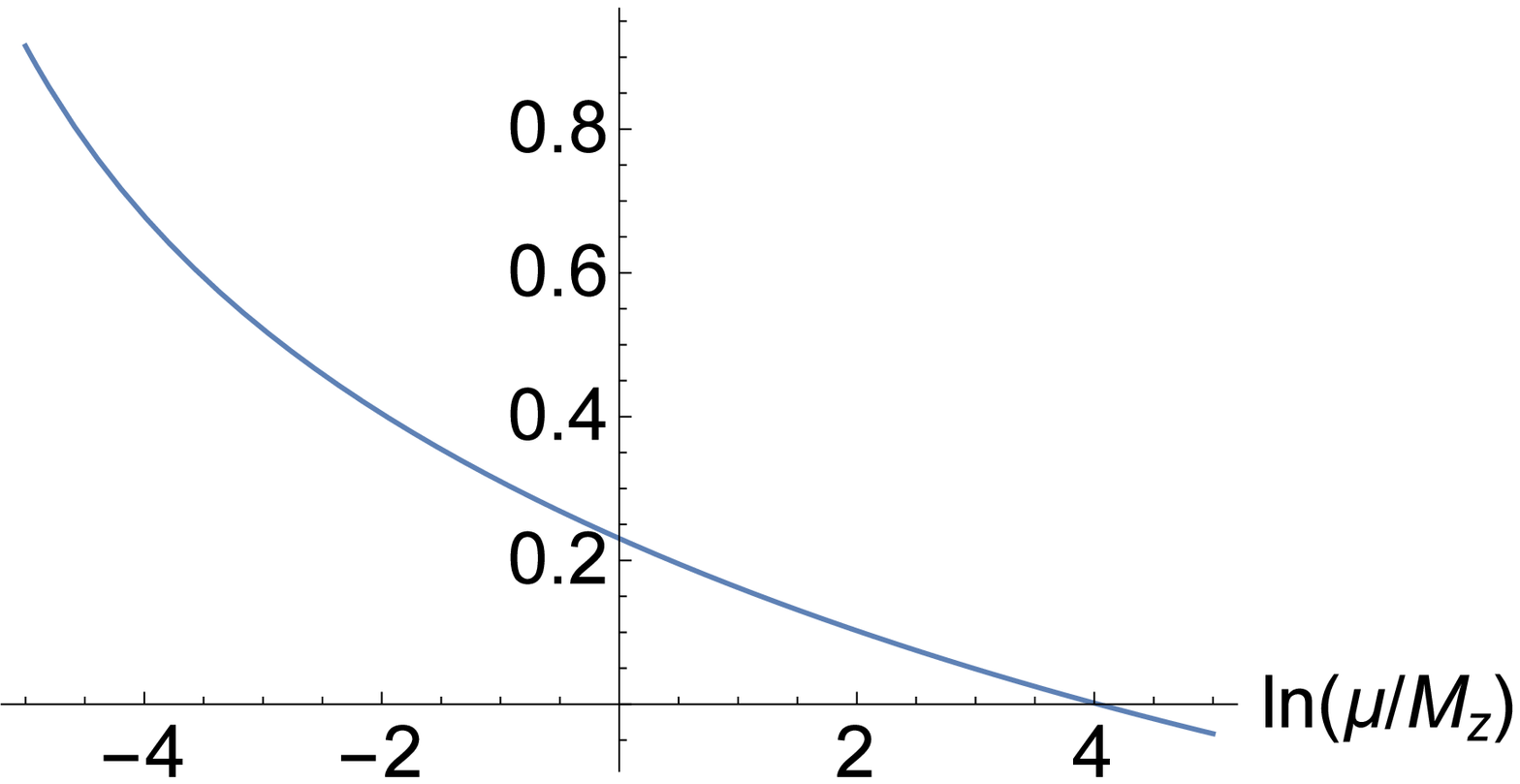}
\put(-120,105){\footnotesize $\tilde\lambda(\mu)$}
\put(-320,105){\footnotesize $\bar g_t(\mu)$}
\caption{Using
experimentally measured SM quantities (including $m_t$ and $m_{_H}$) 
as boundary values, we uniquely solve the RG equations
for the composite Higgs-boson model \cite{bhl1990}, 
we find \cite{xue2013,xue2014} the effective 
top-quark Yukawa coupling $\bar g_t(\mu)$ (left) 
and effective Higgs quartic coupling $\tilde \lambda(\mu)$ (right).  
Note that $\tilde\lambda(\E)=0$ at $\E\approx 5.14~{\rm TeV}$ and $\tilde\lambda(\mu)<0$ 
for $\mu >\E$.
} \label{figyt}
\end{center}
\end{figure}

\section
{\bf  The $\langle\bar t t\rangle$-condensate model}\label{SSBS}
\hskip0.1cm 
In this section, briefly recalling the BHL $\langle\bar t t\rangle$-condensate model 
 \cite{bhl1990} for the full effective Lagrangian of the low-energy SM 
in the scaling region (IR domain) of the IR-fixed point,
we explain why our solution is radically different from the BHL one though the 
same renormalization procedure and RG equations are adopted 
in the IR domain. It is important to compare our solution with the BHL one, as well as 
discuss its difference from the elementary Higgs model. 
  
\subsection
{\bf The scaling region of the IR-stable fixed point and BHL analysis}
\hskip0.1cm 
Using the approach of large $N_c$-expansion with a fixed value 
$GN_c$ to involve the most fermion loops (since each loop provides a factor of $N_c$),
it is shown \cite{bhl1990} that the top-quark channel of 
operators (\ref{bhlx}) undergoes the SSB dynamics in the IR domain 
of IR-stable fixed point $G_c$. 
As a result, the $\Lambda^2$-divergence 
(tadpole-diagram) is removed by the mass gap equation, the 
top-quark channel of four-fermion operator (\ref{bhl}) 
becomes physically relevant and 
renormalizable operators of effective dimension-4. Namely, 
the effective SM Lagrangian with the {\it bilinear} top-quark mass term and 
Yukawa coupling to the composite Higgs boson $H$ at the low-energy 
scale $\mu$ is given by \cite{bhl1990}
\begin{eqnarray}
L &=& L_{\rm kinetic} + g_{t0}(\bar \Psi_L t_RH+ {\rm H.c.})
+\Delta L_{\rm gauge}\nonumber\\ 
&+& Z_H|D_\mu H|^2-m_{_H}^2H^\dagger H
-\frac{\lambda_0}{2}(H^\dagger H)^2,
\label{eff}
\end{eqnarray}
all renormalized quantities received fermion-loop contributions are 
defined with respect to the low-energy scale $\mu$. 
The conventional renormalization $Z_\psi=1$ for fundamental 
fermions and the unconventional wave-function renormalization (form factor)
$\tilde Z_H$ for the composite Higgs boson are adopted
\begin{equation}
\tilde Z_{H}(\mu)=\frac{1}{\bar g^2_t(\mu)},\, \bar g_t(\mu)=\frac{Z_{HY}}{Z_H^{1/2}}g_{t0}; \quad \tilde \lambda(\mu)=\frac{\bar\lambda(\mu)}{\bar g^4_t(\mu)},\,\bar\lambda(\mu)=\frac{Z_{4H}}{Z_H^2}\lambda_0,
\label{boun0}
\end{equation}
where $Z_{HY}$ and $Z_{4H}$ are proper renormalization constants of 
the Yukawa coupling and quartic coupling in Eq.~(\ref{eff}). 
The SSB-generated top-quark mass $m_t(\mu)=\bar g_t^2(\mu)v/\sqrt{2}$. 
The composite Higgs boson is described by its pole-mass 
$m^2_H(\mu)=2\tilde \lambda(\mu) v^2$, form-factor $\tilde Z_H(\mu)=1/\bar g_t^2(\mu)$, 
and effective quartic coupling $\tilde\lambda(\mu)$, provided that 
$\tilde Z_H(\mu)>0$ and $\tilde\lambda(\mu)>0$ are obeyed. After the proper wave-function 
renormalization $\tilde Z_H(\mu)$, the Higgs boson behaves 
as an elementary particle, as long as $\tilde Z_H(\mu)\not=0$ is finite.

In the scaling region of the IR-stable fixed point 
where the SM 
of particle physics is realized, 
the full one-loop RG equations for running couplings $\bar g_t(\mu^2)$ and $\bar \lambda(\mu^2)$
read
\begin{eqnarray}
16\pi^2\frac{d\bar g_t}{dt} &=&\left(\frac{9}{2}\bar g_t^2-8 \bar g^2_3 - \frac{9}{4}\bar g^2_2 -\frac{17}{12}\bar g^2_1 \right)\bar g_t,
\label{reg1}\\
16\pi^2\frac{d\bar \lambda}{dt} &=&12\left[\bar\lambda^2+(\bar g_t^2-A)\bar\lambda + B -\bar g^4_t \right],\quad t=\ln\mu \label{reg2}
\end{eqnarray}
where one can find $A$, $B$ and RG equations for 
running gauge couplings $g^2_{1,2,3}$ in Eqs.~(4.7), (4.8) of 
Ref.~\cite{bhl1990}. The solutions to these ordinary differential equations are uniquely 
determined, once the boundary conditions are fixed.
In 1990, when the top-quark and Higgs masses were unknown, {\it using 
the composite conditions $\tilde Z_H=0$ and $\tilde\lambda=0$ as the 
boundary conditions at the cutoff} $\Lambda$, 
the analysis of the RG equations (\ref{reg1}) and (\ref{reg2}) for $\tilde Z_H(\mu)$ 
and $\tilde\lambda(\mu)$ 
was made to calculate the top-quark and Higgs-boson 
masses by varying the values of cutoff $\Lambda$. It was found that the cutoff $\Lambda$
varies from $10^{4}$ to $10^{19}$ GeV, the obtained top-quark and Higgs-boson masses are
larger than $200$ GeV.

\subsection
{\bf Experimental boundary conditions for RG equations and our analysis}
\hskip0.1cm
We made the same analysis and reproduced the BHL result. However, 
in Refs.~\cite{xue2013,xue2014} we 
further proceed our analysis by using the boundary conditions 
based on the experimental values of top-quark and Higgs-boson masses, $m_t\approx 173$ GeV 
and $m_H\approx 126$ GeV. Namely we adopt these experimental values and 
the mass-shell conditions 
\begin{eqnarray}
m_t(m_t)=\bar g_t^2(m_t)v/\sqrt{2}\approx 173 {\rm GeV},
\quad m_H(m_H)=[2\tilde \lambda(m_H)]^{1/2} v\approx 126 {\rm GeV}
\label{thshell}
\end{eqnarray}
as the boundary conditions of the RG equations (\ref{reg1}) and (\ref{reg2}) to determine 
the solutions for $\tilde Z_H(\mu)$ and $\tilde\lambda(\mu)$
in the IR domain of the energy scale $v=239.5$ GeV, where the low-energy SM physics is achieved with $m_t\approx 173$ GeV and $m_H\approx 126$ GeV.

As a result, we obtained the unique 
solution (see Fig.~\ref{figyt}) for the composite Higgs-boson model (\ref{bhl}) or (\ref{eff}) 
 as well as at the energy scale $\E$  
\begin{eqnarray}
\E \approx 5.1\,\, {\rm TeV},\quad \tilde Z_H \approx 1.26,\quad \tilde\lambda(\E)=0
\label{thvari}
\end{eqnarray}
and effective quartic coupling vanishes $\tilde\lambda(\E)=0$.
As shown in Fig.~\ref{figyt}, or the Fig.~2 in Ref.~\cite{xue2014}, 
our solution
shows the following three important features. 
(I) The squared Higgs-boson mass $m^2_H=2\tilde\lambda(\mu) v^2$ changes its sign at $\mu=\E$, 
indicating the second-order phase transition from the SSB phase 
to the gauge symmetric phase for strong four-fermion coupling \cite{xue1997}. 
(II) The form-factor $\tilde Z_H(\mu)\not=0$ shows that 
the tightly bound composite Higgs particle behaves 
as if an elementary particle for $\mu\leq\E$. Recall that in the BHL analysis 
$\tilde Z_H(\E)=0$ and 
$\tilde\lambda(\E)=0$ are demanded for different $\E$ values. 
(III) The effective form-factor $\tilde Z_H(\E)$ 
of the composite Higgs boson is finite, indicating the formation of massive composite fermions 
$\Psi\sim (H\psi$) in the gauge symmetric phase \cite{xue1997}. 
This critical point of the phase transition could be a ultra-violet (UV) fixed point 
for defining an effective gauge-symmetric field theory for massive composite fermions 
and bosons at TeV scales \cite{xue2014}, and there are some 
possible experimental implications \cite{xue2015}. We do not address 
this issue in this article. 
\subsection
{\bf Compare and contrast}
\hskip0.1cm 
It is important to compare and contrast our study with the BHL one 
\cite{bhl1990}. In both studies, 
the definitions of all physical quantities are identical, the same RG 
equations (\ref{reg1}) and (\ref{reg2}) are used for the running Yukawa 
and quartic couplings as well as gauge couplings. However, 
the different boundary conditions are adopted. 
We impose the infrared boundary conditions (\ref{thshell}) that are known 
nowadays, to uniquely determine 
the solutions of the RG equations, the values of the form-factor
$\tilde Z_{H}(\E)\not=0$ and high-energy scale 
$\E\, [\tilde \lambda(\E)=0]$.  
As shown in Fig.~\ref{figyt}, $\tilde Z_{H}(\mu)=1/\bar g^2_t(\mu)$ [$\tilde \lambda(\mu)$] 
monotonically increases (decreases) as the energy scale $\mu$ increases up to 
$\E$. 
Both experimental $m_t$ and $m_{_H}$ values were 
unknown in the early 1990s, in order to find low-energy values 
$m_t$ and $m_{_H}$ close to the IR-stable fixed point, 
BHL \cite{bhl1990} imposed the 
compositeness conditions $\tilde Z_{H}(\Lambda)=0$ and 
$\tilde \lambda(\Lambda)=0$ for different values of
the high-energy cutoff $\Lambda$ as the boundary condition to 
solve the RG equations. As a result, too large  
$m_t$ and $m_{_H}$ values (Table I in Ref.~\cite{bhl1990}) were obtained, and 
we have reproduced these values. However,
these BHL results are radically different from the present results of 
Eqs.~(\ref{thshell}), (\ref{thvari}) and 
Fig.~\ref{figyt}, showing that the composite Higgs boson actually 
becomes a more and more
tightly bound state, as the energy scale $\mu$ increases, and eventually 
combines with an elementary fermion to form a composite fermion 
in the symmetric phase. This phase transition to 
the gauge symmetric phase is also indicated by $\tilde \lambda(\mu)\rightarrow 0^+$ as $\mu\rightarrow \E+0^-$ at which 
the 1PI vertex function $Z_{4H}$ in Eqs.~(\ref{boun0}) and (\ref{eff})
vanishes.

On the other hand, we compare and contrast our result with the study 
of the fundamental scalar theory for the elementary Higgs particle.
The study  
of the high-order corrections to the RG equations of elementary
Higgs quartic coupling ``$\lambda$'' and measured Higgs mass shows 
that $\lambda(\mu)$ becomes very small and smoothly varies in high energies 
approaching the Planck scale \cite{giudice}. This is a crucial result 
for the elementary Higgs-boson model. 
This result is clearly distinct from the intermediate 
energy scale $\E\sim $ TeV obtained in
the composite Higgs-boson model, where the quadratic 
term $\Lambda^2$ is removed by the mass gap equation of the SSB 
and an ``unconventional'' renormalization for the form factor 
of composite Higgs field is adopted \cite{bhl1990}. Instead in the calculations of 
high-order corrections to the RG equations of the elementary
Higgs quartic coupling ``$\lambda$'', the quadratic term $\Lambda^2$ is 
removed in the $\overline {\rm MS}$ prescription of the conventional renormalization 
for elementary scalar fields. It is worthwhile to mention that in Ref.~\cite{DCW2015}
it is shown in the elementary Higgs-boson model that the quadratic 
term from high-order quantum corrections has a physical impact on the SSB and 
the phase transition to a symmetric phase occurs at the scale of order of TeV.  
Apart from described and discussed above, 
the effective four-fermion interaction theory has 
different dynamics from the fundamental scalar theory for the elementary Higgs particle,
in particular for strong four-fermion coupling $G$, e.g.~the formations of boson and 
fermion bound states \cite{xue1997}. 
Nevertheless, all these studies of either the 
elementary or the composite Higgs-boson model play 
an important role in understanding new physics
beyond the SM for fundamental particles.

\noindent
\section{\bf Origins of explicit symmetry breaking}\label{ESBS}
\hskip0.1cm
Once the top-quark mass is generated via the SSB, 
other fermion $(\nu_\tau, \tau, b)$ masses are generated by the ESB via 
quark-lepton interactions and $W^\pm$-boson vectorlike coupling. 

\subsection{Quark-lepton interactions}

Although the four-fermion operators in Eq.~(\ref{art1}) do not have 
quark-lepton interactions, we consider the following
SM gauge-symmetric four-fermion 
operators that contain quark-lepton interactions \cite{xue1999nu}, 
\begin{eqnarray}
G\left[(\bar\ell^{i}_L\tau_{R})(\bar b^a_{R}\psi_{Lia})
+(\bar\ell^{i}_L\nu^\tau_{R})(\bar t^a_{R}\psi_{Lia})\right] +~~{\rm ``terms"},
\label{bhlql}
\end{eqnarray}
where $\ell^i_L=(\nu^\tau_L,\tau_L)$ and $\psi_{Lia}=(t_{La},b_{La})$
for the third family. The ``terms'' represent   
for the first and second families with substitutions: 
$\tau\rightarrow e,\mu$, $\nu^\tau\rightarrow \nu^e, \nu^\mu$, and 
$t\rightarrow u,c$ and $b\rightarrow d, s$. 
These operators (\ref{bhlql}) should be expected
in the framework of Einstein-Cartan theory and $SO(10)$ unification theory 
\cite{lq1997}.
Once the top quark mass $m_t$ is generated by the SSB, 
the quark-lepton interactions (\ref{bhlql}) introduce the ESB terms
to the SD equations (mass-gap equations) for other fermions. 

\begin{figure}
\begin{center}
\includegraphics[height=1.25in]{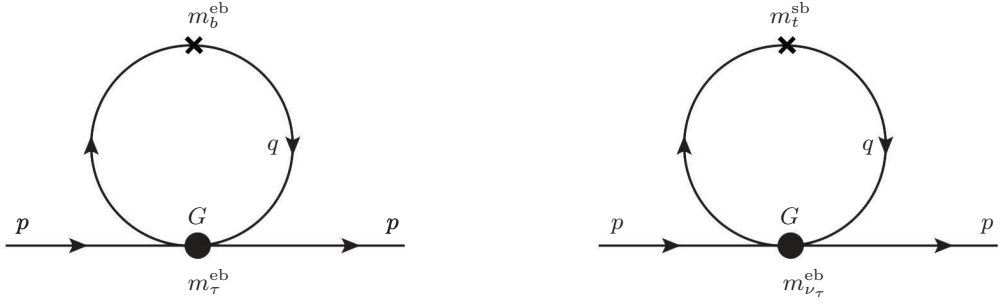}
\includegraphics[height=1.25in]{xue_fermi_5.eps}
\put(-85,45){\footnotesize $q$}
\put(-180,15){\footnotesize $p$}
\put(-40,15){\footnotesize $p$}
\put(-116,17){\footnotesize $G$}
\put(-120,93){\footnotesize $m^{\rm sb}_t$}
\put(-115,-8){\footnotesize $m^{\rm eb}_{\nu_\tau}$}
\put(-310,45){\footnotesize $q$}
\put(-405,15){\footnotesize $p$}
\put(-265,15){\footnotesize $p$}
\put(-340,17){\footnotesize $G$}
\put(-340,93){\footnotesize $m^{\rm eb}_b$}
\put(-340,-8){\footnotesize $m^{\rm eb}_\tau$}
\put(-405,15){\footnotesize $p$}
\put(-265,15){\footnotesize $p$}
\caption{We present the tadpole diagrams of quark-lepton 
interactions (\ref{bhlql}) of the third fermion family, which contribute to quark and 
lepton ESB masses $m^{\rm eb}$ in SD equations (\ref{deq})-(\ref{lb}). 
} \label{figt}
\end{center}
\end{figure}

In order to show these ESB terms, we first approximate the SD equations to be self-consistent
mass gap-equations by neglecting perturbative gauge interactions and using the large $N_c$-expansion to the leading order, as indicated by  Fig.~\ref{figt}.
The quark-lepton interactions (\ref{bhlql}), 
via the tadpole diagrams in Fig.~\ref{figt}, 
contribute to the tau lepton mass 
$m^{\rm eb}_{\tau}$ and tau neutrino mass $m^{\rm eb}_{\tau_\nu}$, provided the 
bottom-quark mass $m^{\rm eb}_{b}$ and top-quark mass $m^{\rm sb}_{t}$ are not zero. 
The latter $m^{\rm sb}_{t}$ is generated by the SSB, see Sec.~\ref{SSBS}. 
The former $m^{\rm eb}_b$ is generated by 
the ESB due to the $W^\pm$-boson vectorlike coupling and top-quark mass $m^{\rm sb}_{t}$, 
see next Sec.~\ref{ESBSw}. The superscript ``sb'' indicates the mass generated by the SSB.
The superscript ``eb'' indicates the mass generated by the ESB. These are bare fermion masses at the energy scale $\E$. 

Corresponding to 
the tadpole diagrams in Fig.~\ref{figt}, 
the mass-gap equations of tau lepton and tau neutrino 
are given by 
\begin{eqnarray}
m^{\rm eb}_{\nu_\tau}&=&2Gm^{\rm sb}_{t} \frac{i}{(2\pi)^4}\int d^4l[l^2-(m^{\rm sb}_{t})^2]^{-1}=(1/N_c)m^{\rm sb}_t.
\label{massql}\\
m^{\rm eb}_{\tau}&=&2Gm^{\rm eb}_{b} \frac{i}{(2\pi)^4}\int d^4l[l^2-(m^{\rm eb}_{b})^2]^{-1}= (1/N_c)m^{\rm eb}_{b},\label{massql'}
\end{eqnarray}
Here we use the self-consistent 
mass-gap equations of the bottom and top quarks [see Eq.~(2.1) and (2.2) in Ref.~\cite{bhl1990}]
\begin{eqnarray}
m^{\rm eb}_{b}&=&2GN_cm^{\rm eb}_{b} \frac{i}{(2\pi)^4}\int d^4l[l^2-(m^{\rm eb}_{b})^2]^{-1},\label{massbt'}\\ 
m^{\rm sb}_{t}&=&2GN_cm^{\rm sb}_{t} \frac{i}{(2\pi)^4}\int d^4l[l^2-(m^{\rm sb}_{t})^2]^{-1},
\label{massbt}
\end{eqnarray}
and the definitions of Dirac quark, lepton and neutrino bare masses in general read 
\begin{eqnarray}
m^{\rm sb}_{\rm qurak}&=&-(1/2N_c)G\langle \bar\psi^a \psi_a\rangle
=-(G/N_c)\langle \bar\psi^a_L \psi_{aR},\rangle\\
m^{\rm sb}_{\rm lepton}&=&-(1/2)G\langle \bar\ell \ell\rangle
=-G\langle \bar\ell_L \ell_R\rangle,
\label{massd}
\end{eqnarray}
and $m^{\rm sb}_{\rm neutrino}=-(1/2)G\langle \bar\ell_\nu \ell_\nu\rangle
=-G\langle \bar\ell_{\nu L} \ell_{\nu R}\rangle$. The notation 
$\langle \cdot\cdot\cdot\rangle$ does 
not represent new SSB condensates, but the 1PI functions of 
fermion mass operator $\bar\psi^a_L \psi_{aR}$, i.e., 
the self-energy functions $\Sigma_f$ that satisfy the self-consistent SD 
equations or mass-gap equations. It is important to note the difference 
that Eq.~(\ref{massbt}) is the 
mass-gap equation for the top-quark mass $m^{\rm sb}_{t}$ generated by the SSB, 
while Eq.~(\ref{massbt}) is just a self-consistent mass-gap equation  
for the bottom-quark mass $m^{\rm eb}_{b}\not=0$, as given by the 
tadpole diagram.
The tau-neutrino mass $m^{\rm eb}_{\nu_\tau}$ and 
tau-lepton mass $m^{\rm eb}_\tau$ are not zero, 
if the top-quark mass $m^{\rm sb}_t$ and bottom-quark mass $m^{\rm eb}_b$ are not zero. 
This is meant to the mass generation of tau neutrino and tau lepton due to
the ESB terms introduced by the quark-lepton interactions (\ref{bhlql}), 
quark masses $m^{\rm sb}_t$ and $m^{\rm eb}_b$. It will be further clarified that these 
ESB terms are actually the inhomogeneous terms in the SD equations, 
which have nontrivial massive solutions without 
extra Goldstone bosons produced. In next section, we are going to show the 
other type of ESB term due to the $W^\pm$-boson vectorlike coupling, 
that is crucial to have the bottom-quark mass $m^{\rm eb}_{b}$ generated by the ESB,
once the top-quark mass $m^{\rm sb}_{t}$ is generated by the SSB.   

\subsection
{\bf $W^\pm$-boson coupling to right-handed fermions}\label{ESBSw}
\hskip0.1cm
In addition to the ESB terms due to quark-lepton interactions, 
the effective vertex of $W^\pm$-boson coupling to right-handed fermions at 
the energy scale $\E$ also introduces the ESB terms to the 
Schwinger-Dyson equations for other fermions, once 
the top-quark mass $m_t$ is generated by the SSB. We study this effective 
vertex in this section.

\begin{figure}
\begin{center}
\includegraphics[height=2.00in]{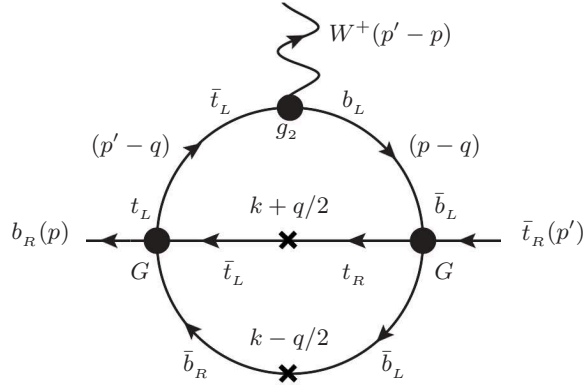}
\put(-65,130){\footnotesize $W^+(p'-p)$}
\put(-85,95){\footnotesize $g_{_2}$}
\put(-155,88){\footnotesize $(p'-q)$}
\put(-35,88){\footnotesize $(p-q)$}
\put(-110,105){\footnotesize $\bar t_{_L}$}
\put(-60,105){\footnotesize $b_{_L}$}
\put(-140,65){\footnotesize $t_{_L}$}
\put(-25,65){\footnotesize $\bar b_{_L}$}
\put(-120,5){\footnotesize $\bar b_{_R}$}
\put(-45,5){\footnotesize $\bar b_{_L}$}
\put(-60,40){\footnotesize $t_{_R}$}
\put(-105,40){\footnotesize $\bar t_{_L}$}
\put(-95,65){\footnotesize $k+q/2$}
\put(-95,15){\footnotesize $k-q/2$}
\put(-185,55){\footnotesize $b_{_R}(p)$}
\put(8,55){\footnotesize $\bar t_{_R}(p')$}
\put(-140,40){\footnotesize $G$}
\put(-25,40){\footnotesize $G$}
\caption{We adopt the third quark family $(t,b)$ as an example to illustrate 
the 1PI vertex function of $W^\pm_\mu$-boson coupling to 
right-handed Dirac fermions induced by four-fermion operators (\ref{bhlx}).
$\ell$, $p$ and $p'$ are external momenta, $q$ and $k$ are internal momenta 
integrated up to the energy scale $\E$.
The cross ``$\times$'' represents self-energy functions of Dirac fermions,
which are the eigenstates of mass operator. A CKM matrix $U_{ij}$ 
associates to the $W$-boson coupling $g_2$.}\label{figi}
\end{center}
\end{figure}

In the low-energy SM obeying chiral gauge symmetries, the parity symmetry is violated,
in particular, the $W^\pm$-boson couples only to the left-handed fermions, i.e., 
$i(g_2/\sqrt{2})\gamma_\mu P_L$. 
In order to show that the 
four-fermion operators (\ref{art1}) induce a 1PI 
vertex function of $W^\pm_\mu$-boson coupling to
the right-handed fermions, 
see Fig.~\ref{figi}, 
we take the third quark family in Eq.~(\ref{bhlx}) 
\begin{eqnarray}
L 
&=& L_{\rm kinetic} + G(\bar\psi^{ia}_Lt_{Ra})(\bar t^b_{R}\psi_{Lib})
+ G(\bar\psi^{ia}_Lb_{Ra})(\bar b^b_{R}\psi_{Lib}), 
\label{bhlx_1}
\end{eqnarray}
as an example for calculations.
\comment{where $a,b$ and $i,j$ are the color and flavor indexes 
of the top and bottom quarks, the $SU_L(2)$ doublet 
$\psi^{ia}_L=(t^{a}_L,b^{a}_L)$ 
and the singlet $\psi^{a}_R=t^{a}_R,b^{a}_R$ are the eigenstates 
of the electroweak interaction, and additional terms for 
the first and second quark families can be obtained 
by substituting $t\rightarrow u,c$ and $b\rightarrow d,s$ \cite{xue2014}.}
The leading contribution to the nontrivial 1PI vertex function is given by
\begin{eqnarray}
& &G^2(\bar\psi^{a'}_Lb_{Ra'})(\bar b^{b'}_{R}\psi_{Lb'})
(\bar\psi^{a}_Lt_{Ra})(\bar t^b_{R}\psi_{Lb})
\Big\{\frac{ig_2}{\sqrt{2}}\bar t_{Lc}(\gamma^\mu P_L) b^c_{L}W^+_\mu\Big\}
\nonumber\\
&=& i\frac{g_2}{\sqrt{2}}G^2(\bar t^{a'}_Lb_{Ra'})(\bar b^{b'}_{R}t_{Lb'})
(\bar b^{a}_Lt_{Ra})(\bar t^b_{R}b_{Lb})
\Big\{\bar t_{Lc}(\gamma^\mu P_L) b^c_{L}W^+_\mu\Big\}\nonumber\\
&=& i\frac{g_2}{\sqrt{2}}G^2b_{Ra'}\left\{[t_{Ra}\bar t^{a'}_L][b_{Lb}\bar b^{b'}_{R}]
[t_{Lb'}\bar t_{Lc}][\gamma^\mu P_L] [b^c_{L}\bar b^{a}_L]\right\} \bar t^b_{R} W^+_\mu\label{2p0}\\
&=& i\frac{g_2}{\sqrt{2}}G^2N_cb^\eta_{R}\left\{[t_{R}\bar t_L]^{\lambda\eta}[b_{L}\bar b_{R}]^{\alpha\beta}
[t_{L}\bar t_{L}]^{\beta\delta}[\gamma^\mu P_L]^{\delta\sigma} 
[b_{L}\bar b_L]^{\sigma\lambda}\right\} \bar t^\alpha_{R} W^+_\mu\label{2p1}\\
&\Rightarrow& i\frac{g_2}{\sqrt{2}}
b_{R}^{\eta} [\Gamma^W_\mu(p',p)]^{\eta\alpha}
~ \bar t^\alpha_{R}~W^+_\mu(p'-p)
\label{2p}
\end{eqnarray}
where two fields in brackets $[\cdot\cdot\cdot]$ in the line 
(\ref{2p0}) mean the contraction of them, 
as shown in Fig.~\ref{figi}, the color degrees ($N_c$) of freedom have been 
summed and spinor indexes are explicitly shown 
in the line (\ref{2p1}). $\Gamma^W_\mu(p',p)$ represents the effective  
vertex function of $W^\pm$-boson coupling to the right-handed fermions $t_R$ and $b_R$, 
\begin{eqnarray}
[\Gamma^W_\mu(p',p)]^{\eta\alpha} &=&
\frac{g_2}{\sqrt{2}}G^2 N_c
\!\int^\E_{k,q} 
\left[\frac{\gamma\cdot(p'\!+\!q)}{(p'+q)^2}\right]^{\beta\delta}_{\rm t-quark}
(\gamma^\mu P_L)^{\delta\sigma}\left[\frac{\gamma\cdot (p\!-\!q)}{(p-q)^2}\right]^{\sigma\lambda}_{\rm b-quark} 
\nonumber\\
&\times &\left[\frac{\gamma\cdot(k+q/2)-m_t}{(k+q/2)^2-m_t^2}
\right]^{\lambda\eta}\left[\frac{\gamma\cdot (k-q/2)-m_b}{(k-q/2)^2-m_b^2}\right]^{\alpha\beta}.
\label{az1}
\end{eqnarray}
Based on the Lorentz invariance, the 
1PI vertex can be written as
\begin{eqnarray}
\Gamma_\mu^W(p,p')&=& i\frac{g_2}{\sqrt{2}}\gamma_\mu P_R\,\Gamma^W(p,p'),
\label{vr}
\end{eqnarray}
where $\Gamma^W(p,p')$ is 
the dimensionless Lorentz scalar.
Beside, this vertex function (\ref{vr}) remains the same for exchanging $b$ and $t$. 
The same calculations can be done by replacing $t\rightarrow u,c$  
and $b\rightarrow d,s$, as well as $t\rightarrow \nu_e,\nu_\mu,\nu_\tau$  
and $b\rightarrow e,\mu,\tau$. 

As shown in Fig.~\ref{figi} and Eq.~(\ref{vr}), 
the two-loop calculation to obtain 
the finite part of the dimensionless Lorentz scalar 
$\Gamma^W(p,p')$ is not an easy task. 
Nevertheless we can preliminarily infer its behavior as a function 
of energy $p$ and $p'$. For the 
case $p\ll m_t$ and $p'\ll m_b$, the vertex function  
$\Gamma^W(p,p')\propto (G\E^2)^2(m_t/\E)^2(m_b/\E)^2\ll 1$ vanishes 
in the IR domain of IR-fixed point of weak four-fermion coupling \cite{bhl1990}, 
where the SM with parity-violating
gauge couplings of $W^\pm$ and $Z^0$ bosons are realized. 
For the case $p\gg m_t$ and $p'\gg m_b$, 
$\Gamma^W(p,p')\propto (G\E^2)^2(p'/\E)^2(p/\E)^2$ increases as $p$ and $p'$ increase. 
In addition the four-fermion coupling $G$ increases 
its strength as energy scale increases, i.e., 
the $\beta(G)$-function is positive \cite{xue2014}. 
This implies that in high energies $(p/\E)^2\lesssim 1$ and/or $(p'/\E)^2\lesssim 1$, 
the vertex function $\Gamma^W(p,p')\equiv\Gamma^W[(p/\E)^2,(p'/\E)^2]$ 
does not vanish and the $W^\pm$-boson coupling to fermions is no longer 
purely left handed. 
\comment{, deviating from the SM. 
We expect that the vertex function 
$\Gamma^W(p,p')$ should approach to one, when energy-momenta 
$p$ and/or $p'$ approach to the energy threshold $\E\gtrsim 5\,$ TeV, 
since it is the approximate energy scale of transition 
to the symmetric phase of preserving parity symmetry 
by massive composite Dirac 
fermions \cite{xue1997} in the domain of UV fixed point \cite{xue2014}.
} 
On the other hand, at high-energy scale, 
the dependence of the vertex function 
$\Gamma^W(p,p')$ on fermion masses is negligible, 
and $\Gamma^W(p,p')$ is approximately universal 
for all quarks and leptons.  

\noindent
\section{\bf  Schwinger-Dyson equations}\label{SDS}
\begin{figure}
\begin{center}
\includegraphics[height=1.50in]{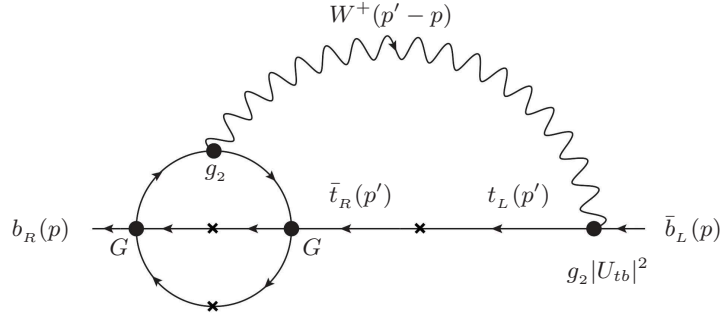}
\put(-130,110){\footnotesize $W^+(p'-p)$}
\put(-177,52){\footnotesize $g_{_2}$}
\put(-40,13){\footnotesize $g_{_2}|U_{tb}|^2$}
\put(-130,42){\footnotesize $\bar t_{_R}(p')$}
\put(-70,42){\footnotesize $t_{_L}(p')$}
\put(-250,29){\footnotesize $b_{_R}(p)$}
\put(-3,29){\footnotesize $\bar b_{_L}(p)$}
\put(-140,22){\footnotesize $G$}
\put(-213,22){\footnotesize $G$}
\caption{We adopt the third quark family $(t,b)$
as an example to illustrate the general $W^\pm$-boson contribution 
to the fermion self-energy function $\Sigma(p)$. The bottom-quark 
self-energy function $\Sigma_b(p)$ is related to the top-quark one 
$\Sigma_t(p')$.} \label{figs}
\end{center}
\end{figure}

\hskip0.1cm
After discussing the simplest mass-gap equations (\ref{massql}) and (\ref{massbt}) and the 
effective $W^\pm$-boson coupling vertex (\ref{vr}), we turn to the SD equations for fermion 
self-energy functions by taking gauge interactions into account. 
It is known that in the SM the $W^\pm$ boson does not contribute 
to the SD equations for fermion self-energy functions $\Sigma_f$. 
However, due to the nontrivial vertex 
function (\ref{vr}), the $W^\pm$ gauge boson has the contribution, 
as shown in Fig.~\ref{figs}, to 
SD equations at high energies.
This contribution not only introduces an explicit symmetry breaking term,
but also mixes up SD equations 
for self-energy functions of different fermion fields 
via the CKM mixing matrix \cite{xue1999nu,xue1997mx}.

In the vectorlike gauge theory, 
SD equations for fermion
self-energy functions were intensively studied in Ref.~\cite{kogut}.  
In the Landau gauge, SD equations for $t$ and $b$ quarks are given by
\begin{eqnarray}
\Sigma_t(p)&=& m^{\rm sb}_t +
3\int_{p'}\frac{V_{2/3}(p,p')}{(p-p')^2}{\Sigma_t(p')\over p'^2
+\Sigma_t(p')}\label{23}\\
\Sigma_b(p)&=& m^{\rm eb}_b + 3
\int_{p'}\frac{V_{-1/3}(p,p')}{(p-p')^2}{\Sigma_b(p')\over p'^2
+\Sigma_b(p')}\label{13}
\end{eqnarray}
where the integration $\int_{p'}\equiv \int d^4p'/(2\pi)^4$ is 
up to the cutoff $\E$,
$V_{2/3}(p,p')$ and $V_{-1/3}(p,p')$ are the vertex functions of
vectorlike gauge theories. In Eq.~(\ref{23}), the bare mass term $m^{\rm sb}_t$  
comes from the SSB, see the simplest mass-gap equation (\ref{massbt}) and discussions 
in Sec.~\ref{SSBS}. 
Instead, the bare mass term
$m^{\rm eb}_b$ in Eq.~(\ref{13}) comes from the ESB terms due to the
effective $W^\pm$-boson coupling vertex (\ref{vr}) and the self-consistent
mass-gap equation (\ref{massbt'}).  
We neglect corrections to vertex functions of vectorlike gauge 
interactions, for example, $V_{{2/3}}=(2e/3)^2$ and $V_{-1/3}=(e/3)^2$ 
in the QED case. Since the vertex function $\Gamma^W(p,p')$ 
in Eq.~(\ref{vr}) does not vanish only for high energies, 
we approximately treat it as a boundary value at the scale $\E$
\begin{eqnarray}
\alpha_w=\alpha_2(\E)(\gamma_w/\alpha_c\sqrt{2}),\quad \alpha_2(\E)=g_2^2(\E)/4\pi,\quad
\gamma_w=\Gamma^W(p,p')|_{p, p'\rightarrow \E},
\label{alphaw}
\end{eqnarray}
where $\alpha_c=\pi/3$. This means that the contributions of Fig.~\ref{figs} 
are approximately boundary terms in the integral SD equations 
(\ref{23}) and (\ref{13}), see the following equations (\ref{deq})-(\ref{boundaryb}). 
Thus we neglect possible right-handed couplings of the
would-be Nambu-Goldstone bosons in the Landau gauge, which could have effects 
on the explicit gauge symmetry breaking. In the future we will study these effects
in some more detail.

Following the approach of Ref.~\cite{kogut}, we convert  
Eqs.~(\ref{23}) and (\ref{13}) to the following boundary value 
problems ($x=p^2$, $\alpha=e^2/4\pi$):
\begin{eqnarray}
&&{d\over dx}\left(x^2\Sigma_t'(x)\right)+{(2/3)^2\alpha\over \alpha_c}{x\Sigma_t(x)\over x+
\Sigma_t^2(x)}=0,\label{deq}\\
&& \E^2\Sigma_t'(\E^2)+\Sigma_t(\E^2)- m^{\rm sb}_t=\alpha_w|U_{tb}|^2\Sigma_b(\E^2)+m^{\rm sb}_t,
\label{boundary}
\end{eqnarray}
and 
\begin{eqnarray}
&&{d\over dx}\left(x^2\Sigma_b'(x)\right)+{(1/3)^2\alpha\over\alpha_c}{x\Sigma_b(x)\over x+
\Sigma^2_b(x)}=0,\label{deqb}\\
&&\E^2\Sigma_b'(\E^2)+\Sigma_b(\E^2)=\alpha_w|U_{bt}|^2\Sigma_t(\E^2)
+m^{\rm eb}_b.
\label{boundaryb}
\end{eqnarray}
The boundary conditions (\ref{boundary})
and (\ref{boundaryb}) are actually the mass-gap equations of $t$ and $b$ quarks 
at the scale $\E$, the $\alpha_w$-terms come from the contribution of Fig.~\ref{figs},
$m^{\rm eb}_b$- and $m^{\rm sb}_t$-terms come from Eqs.~(\ref{massbt'}) and (\ref{massbt}).
Analogously, we obtain the following boundary value problem for the $\nu_\tau$ and $\tau$ leptons:
\begin{eqnarray}
{d\over dx}\left(x^2\Sigma_{\nu_\tau}'(x)\right)&=&0,\label{ldeq}\\
\E^2\Sigma_{\nu_\tau}'(\E^2)+\Sigma_{\nu_\tau}(\E^2)
&=&\alpha_w|U_{\nu_\tau\tau}|^2\Sigma_\tau(\E^2)+m^{\rm eb}_{\nu_\tau},
\label{nub}
\end{eqnarray}
and
\begin{eqnarray}
{d\over dx}\left(x^2\Sigma_\tau'(x)\right)
+{\alpha\over\alpha_c}{x\Sigma_\tau(x)\over x+
\Sigma^2_\tau(x)}&=&0, \label{ldeqb}\\
\E^2\Sigma_\tau'(\E^2)+\Sigma_\tau(\E^2)
&=&\alpha_w|U_{\tau\nu_\tau}|^2\Sigma_{\nu_\tau}(x)+m^{\rm eb}_\tau,
\label{lb}
\end{eqnarray}
where $U_{\tau\nu_\tau}$ is the  element of PMNS mixing matrix. 
The boundary conditions (\ref{nub}) and (\ref{lb}) 
are actually the mass-gap equations of $\nu_\tau$ and $\tau$ leptons 
at the scale $\E$, the $\alpha_w$-terms come from the contribution of Fig.~\ref{figs},
$m^{\rm eb}_{\tau}$- and $m^{\rm eb}_{\nu_\tau}$-terms come 
from Eqs.~(\ref{massql'}) and (\ref{massql}).
In the rhs of self-consistent
mass-gap equations (\ref{boundary}), (\ref{boundaryb}), (\ref{nub}) and (\ref{lb}), only 
the top-quark mass term $m^{\rm sb}_t$ is due to the SSB, see Eq.~(\ref{massql}) 
and Sec.~\ref{SSBS}, whereas
the $\alpha_w$-terms are the ESB terms 
due to the effective vertex (\ref{vr}), see Fig.~\ref{figs}, 
whereas the $m^{\rm eb}_{b}$-, $m^{\rm eb}_{\tau}$- 
and $m^{\rm eb}_{\nu_\tau}$-terms are ESB terms, satisfying the 
self-consistent mass-gap equations, e.g., Eqs.~(\ref{massql'})-(\ref{massbt}) 
due to the quark-lepton interactions (\ref{bhlql}), four-fermion interactions 
(\ref{bhlx}) and (\ref{bhlxl}). All ESB terms are functions of 
the top-quark mass term $m^{\rm sb}_t$, which is the unique origin of the ESB terms. 
The SD equations (\ref{deq}(-(\ref{lb})
are coupled, become inhomogeneous and we try to find the nontrivial massive solutions 
for the bottom quark, tau lepton and tau neutrino. 

Suppose that the SSB for the
top-quark mass does not occur ($m^{\rm sb}_t=0$), fermion bare masses $m^{\rm eb}$ are zero
and the $W$-boson contribution vanishes ($\alpha_w=0$). 
In this case the SD equations (\ref{deq}-\ref{lb}) are homogenous.
It was established \cite{kogut} that only trivial solutions $\Sigma_f(p)=0$ 
to homogeneous SD equations exist in the weak
coupling phase $\alpha<\alpha_c$, however, 
inhomogeneous SD equations have nontrivial solutions
\begin{equation}
\Sigma_f(p)\propto m_f\big({p^2\over m^2_f}\big)^{\gamma},\quad m_f\leq p\leq \E,
\label{exp}
\end{equation}
where the factor $\big({p^2\over m^2_f}\big)^{\gamma}$ comes from the corrections of 
perturbative gauge interactions and 
$\gamma\ll 1$ is the anomalous dimension of fermion mass operators. 
Actually, when  $x\gg \Sigma_f(x)$ and the nonlinearity in SD equations is neglected,  
Eqs.~(\ref{deq}), (\ref{deqb}) and (\ref{ldeqb}) admit 
the solution \cite{xue2000fi}, 
\begin{eqnarray}
\Sigma_f(x) &\propto & \frac{m_f^2}{\mu}\sinh \Big[\frac{1}{2}\sqrt{1-\frac{\alpha_f}{\alpha_c}}
\ln\Big(\frac{\mu^2}{m_f^2}\Big)\Big]
\propto  m_f\Big(\frac{\mu^2}{m_f^2}\Big)^{(\alpha_f/4\alpha_c)},
\label{msol}
\end{eqnarray}
``$f=\tau,b,t$'', $x=p^2\propto\mu^2$ and $\mu$ is the infrared scale. 
In Eqs.~(\ref{exp}) 
and (\ref{msol}), the infrared mass scales $m_f=m_f(\mu)$ are proportional to the 
inhomogeneous terms attributed to the ESB terms, 
which are in our scenario $\alpha_w$-terms and mass terms 
$m^{\rm eb}_f$. Equation (\ref{ldeq}) 
for $\nu_\tau$-neutrino ($\alpha_f=0$) admits the solution 
$\Sigma_{\nu_\tau}(x)=m_{\nu_\tau}(\mu)$ 
that is related to the inhomogeneous term of the ESB at the infrared scale $\mu$.

\noindent
\section{\bf The masses of third fermion family}
\hskip0.1cm
First we try to find the massive 
solutions ($m_{\nu_\tau},m_\tau, m_b,m_t$) to the mass-gap equations
Eqs.~(\ref{boundary}), (\ref{boundaryb}), (\ref{nub}) and (\ref{lb}) at the 
energy scale $\E$. In these equations, the term $\E^2\Sigma_i'(\E^2)$ can be neglected,
since $\E^2\Sigma_i'(\E^2)=\gamma\Sigma_i(\E^2)\ll \Sigma_i(\E^2)$, where 
Eq.~(\ref{exp}) is used. 
We define
the bare masses 
$\Sigma_t(\E^2)\equiv m^0_t \approx m^{\rm sb}_t$,  
$\Sigma_b(\E^2)\equiv m^0_b\approx m^{\rm eb}_b$, 
$\Sigma_\tau(\E^2)\equiv m^0_\tau\approx m^{\rm eb}_\tau$, and
$\Sigma_{\nu_\tau}(\E^2)\equiv m^0_{\nu_\tau}\approx m^{\rm eb}_{\nu_\tau}$.
As a result, we approximately obtain
\begin{eqnarray}
m^0_{\nu_\tau}&\approx&\alpha_w|U^\ell_{\tau\nu_\tau}|^2m_\tau^0
+m^0_t/N_c\lesssim m^0_t/N_c
\label{gapnu}\\
m^0_\tau&\approx&\alpha_w|U^\ell_{\tau\nu_\tau}|^2m^0_{\nu_\tau}
+ m^0_b/N_c\lesssim 2 m^0_b/N_c
\label{gapta}\\
m^0_t&\approx&\alpha_w|U_{tb}|^2m_b^0
+ N_c m^0_{\nu_\tau} + m^{\rm sb}_t\approx m^{\rm sb}_t,
\label{gapt}\\
m^0_b&\approx&\alpha_w|U_{bt}|^2m_t^0
+ N_c m^0_\tau
\approx \alpha_w m_t^0
\label{gapb}
\end{eqnarray}
where $|U_{tb}|\approx 1.03$ \cite{pdg2012} and 
$|U^\ell_{\tau\nu_\tau}|\lesssim 1$. Equations (\ref{gapnu}) and (\ref{gapt}) 
are used in the last inequality of Eq.~(\ref{gapta}). Equations 
(\ref{gapnu})-(\ref{gapb}) show that at the energy scale $\E$, 
the ESB masses $m^0_{\nu_\tau}$, 
$m^0_\tau$ and $m^0_b$ are originated from the SSB mass $m^0_t$. 
The last step in Eqs.~(\ref{gapnu}-\ref{gapb})
shows the dominate contributions: (i) the $\nu_\tau$-neutrino 
acquires its mass $m^0_{\nu_\tau}$ from the $t$-quark mass $m^0_t$ via 
the quark-lepton interaction (\ref{bhlql}), 
(ii) the $b$-quark 
acquires its mass $m^0_b$ from the $t$-quark mass $m^0_t$ via the CKM mixing, (iii) 
the $\tau$-lepton 
acquires its mass $m^0_\tau$ from the bottom-quark mass $m^0_b$ via the quark-lepton 
interaction (\ref{bhlql})
and $\tau$-neutrino mass $m^0_{\nu_\tau}$ via the PMNS mixing. In this article, we only indicate that the tau neutrino Dirac mass relates to the top-quark mass without further 
discussions, since the problem of neutrinos masses is complex for their Dirac or Majorana feature. We will discuss in the future that
the Dirac mass $m^0_{\nu_\tau}$ of the left-handed neutrino $\nu_L$ and right-handed sterile 
neutrino $\nu_R$, as well as the large Majorana mass of Majorana neutrino 
$(\nu_{R}+\nu_{R}^{c}$) generated by the four-fermion operator  
$G(\bar\nu^{\ell\, c}_R\nu^{\ell\,}_{R})(\bar \nu^{\ell}_{R}\nu^{\ell c}_{R})$ 
in Eq.~(\ref{art1}), in order to see if 
the smallness of gauged Majorana neutrino masses is consistent with experimental data. 

In order to find the fermion self-energy function 
(\ref{exp}) or (\ref{msol})
at the infrared mass scale $\mu$, we need to solve 
the inhomogeneous SD equations (\ref{deq}), (\ref{deqb}) and (\ref{ldeqb}) 
with the boundary values (\ref{gapnu})-(\ref{gapb}). This is still a difficult task.
To the leading order, neglecting the corrections from perturbative gauge interactions,
we use Eqs.~(\ref{gapta}) and (\ref{gapb}) to approximately obtain  
the bottom quark and $\tau$-lepton masses at the infrared mass scale $\mu$,
\begin{eqnarray}
\quad m_\tau(\mu)\approx 2 N_c^{-1} m_b(\mu)
,\quad  m_b(\mu)\approx  \alpha_w  m_t(\mu),
\label{rbm}
\end{eqnarray}
in terms of the top-quark mass $m_t(\mu)=\bar g_t(\mu) v/\sqrt{2}$, and we 
define the bottom-quark and tau-lepton Yukawa couplings 
\begin{eqnarray}
m_b(\mu)=\bar g_b(\mu)v/\sqrt{2},\quad m_\tau(\mu)=\bar g_\tau(\mu)v/\sqrt{2},
\label{yukawa3}
\end{eqnarray}
which are obviously related to the top-quark Yukawa coupling $\bar g_t(\mu)$.
In Fig.~\ref{figyl}, the Yukawa couplings $\bar g_b(\mu)$ and $\bar g_\tau(\mu)$ are plotted.
Comparing them with the Yukawa coupling $\bar g_t(\mu)$ in Fig.~\ref{figyt}, one finds the hierarchy pattern of the Yukawa couplings in the third fermion family.

\comment{and Yukawa couplings 
\begin{eqnarray}
\bar g_{_{\nu_\tau}}(\mu)&\approx&N_c^{-1} \bar g_t(\mu)
\label{ntaum}\\
\bar g_\tau(\mu)&\approx& 2 N_c^{-1} \bar g_b(\mu)
\label{taum}\\
\bar g_t(\mu)&\approx& \bar g_t(\mu) ,
\label{tm}\\
\bar g_b(\mu)&\approx&  \alpha_w\bar g_t(\mu). 
\label{bm}
\end{eqnarray}
}
Using the top-quark mass-shell condition and experimental values of 
top- and bottom-quark masses: 
$m_t=\bar g_t(m_t)v/\sqrt{2}\approx 173$ GeV and  
$m_b
\approx 4.2$ GeV, as well as 
the $SU(2)$ gauge-coupling value $g^2_2(\E)\approx 0.42$,
we numerically obtain  the $\alpha_w$- and $\gamma_w$-values at the energy scale $\E$,
\begin{eqnarray}
\alpha_w \approx (2/N_c)\Big(\frac{m_b}{m_t}\Big) 
\Big[\frac{\bar g_t(m_t)}{\bar g_t(m_b)}\Big]= 1.9\times 10^{-2} (2/N_c)
\label{alphav}
\end{eqnarray}
and $\gamma_w\approx 0.85(2/N_c) \sim {\mathcal O}(1)$ in Eq.~(\ref{alphaw}). Note that
$\bar g_t(\mu)$ in Fig.~\ref{figyt} has received
the contributions from the renormalized gauge couplings
$g_{1,2,3}(\mu)$ of the SM, see Sec.~\ref{SSBS}. Thus we approximately 
determine the finite part of vertex function 
$\Gamma^W(p,p')$ (\ref{vr}), since we have not calculated it. 

\comment{Instead of solving the inhomogeneous SD Eqs.~(\ref{deq},\ref{deqb}) and (\ref{ldeqb}), to
find how the infrared mass scales 
in Eq.~(\ref{exp}) or (\ref{msol}) depend on the running gauge  
couplings $\bar g_{1,2,3}(\mu)\equiv g_{1,2,3}(\mu)/g_{1,2,3}(\E)$ of the SM,} 

In the determination of the $\tau$-lepton mass, we
take into account the corrections of perturbative gauge interactions 
by adopting the RG solutions for fermion masses \cite{zli} 
(the number of quark flavors $N_F=6$), 
\begin{eqnarray}
m_t(\mu)/m^0_t \approx 
[\bar g_3(\mu)]^{8/7}
[\bar g_1(\mu)]^{-1/10},\quad
m_b(\mu)/m^0_b \approx [\bar g_3(\mu)]^{8/7}[\bar g_1(\mu)]^{1/20},
\label{brg}
\end{eqnarray}
and $
m_\tau(\mu)/m^0_\tau \approx 
[\bar g_1(\mu)]^{-9/20}$. 
\comment{ 
The physical solution to Eq.~(\ref{ldeq}) and boundary condition (\ref{nub}) or 
(\ref{gapnu}) is $\Sigma_{\ell_\nu}(\mu)\sim \mu$ 
(see Ref.~\cite{xue2000fi}), where $\mu$ is related to the explicit symmetry breaking terms and 
their RG flows. Namely, the neutrino mass $m_{\nu_\tau}(\mu)$ and Yukawa coupling
$\bar g_{\nu_\tau}(\mu)$ are mainly related to the top-quark mass $m_t(\mu)$ and Yukawa coupling
$\bar g_t(\mu)$.}   
We use Eqs.~(\ref{rbm}), (\ref{yukawa3}), (\ref{brg}) 
and the $\tau$-lepton mass-shell condition 
\begin{eqnarray}
m_\tau=\bar g_\tau(m_\tau) [\bar g_1(m_\tau)]^{-1/2} [\bar g_3(m_\tau)]^{-8/7}
v/\sqrt{2}
\label{masstau}
\end{eqnarray}
to uniquely determine the $\tau$-lepton mass
$m_\tau\approx 1.59$ GeV. This is a qualitative result, since we have only considered in 
the inhomogeneous SD equation 
the possible dominate contributions to the tau lepton mass and neglected other contributions,
e.g., the fermion-family mixing. 
Nevertheless, the result is in a qualitative agreement with the experimental value 
$m_\tau\approx 1.78$ GeV, and consistently 
shows the hierarchy spectrum of fermion masses $m_t$, $m_b$ and $m_\tau$ 
in the third family. It should be emphasized that these qualitative results 
cannot to be quantitatively compared with the SM precision tests. 
The quantitative study is a difficult and challenging task and 
one will probably be able to carry it out by using a numerical approach in the future.


\noindent
\section{\bf A brief conclusion and some remarks}
\hskip0.1cm  
Our goal in this article is to present a possible scenario and understanding of 
the origins and hierarchy spectrum of fermion masses in the third family of the SM.
We obtain the fermion masses, Yukawa couplings and their relations, 
as well as the energy scale $\E\approx 5.1$ TeV at which the second-order phase transition
occurs from the SSB phase to the gauge symmetric phase. All these results are 
preliminarily qualitative, and they should receive the high-order corrections and 
some nonperturbative contributions. Nevertheless, these results may give us some insight into the 
long-standing problem of fermion-mass origin and hierarchy.
We will present the similar study including 
the first and second families, as well as neutrinos 
by taking into account the fermion-family mixing to understand 
the hierarchy spectrum of SM fermions: 
from top quark to electron neutrino \cite{xue2016mass}.
It is much more complicated to solve the SD equations for the SM fermion masses, 
however, the basic scenario is simply the same. Due to 
the ground-state (vacuum) alignment of the effective theory of relevant operators, 
the top-quark mass is generated by the SSB, and other fermion masses are originated from 
the ESB terms, which are induced by 
the top-quark mass via the fermion-family mixing, quark-lepton interactions and 
vectorlike $W^\pm$-boson coupling at high energies. As a consequence,
fermion Yukawa couplings are functions of the top-quark Yukawa coupling. 
   
\begin{figure}
\begin{center}
\includegraphics[height=1.40in]{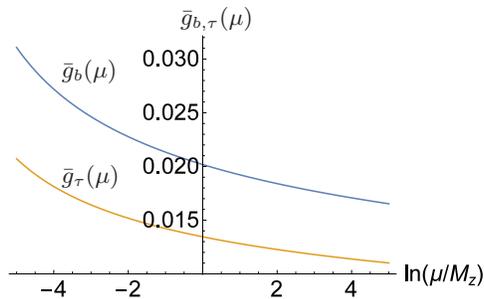}
\put(-115,105){\footnotesize $\bar g_{b,\tau}(\mu)$}
\put(-160,45){\footnotesize $\bar g_\tau(\mu)$}
\put(-160,85){\footnotesize $\bar g_b(\mu)$}
\caption{We plot the Yukawa couplings $\bar g_b(\mu)$ and $\bar g_\tau(\mu)$ from 
$\mu\ge 0.5$ GeV to $\E\approx 5$ TeV.} \label{figyl}
\end{center}
\end{figure}

In this article, the top-quark Yukawa coupling $\bar g_{t}(\mu)$ in fact relates to 
the nonvanishing form-factor $\tilde Z_H(\mu)$ of composite 
Higgs boson, see Eq.~(\ref{boun0}). Both of them, as shown in Fig.~\ref{figyt},
are of the order of unity and slowly vary from $1$ GeV to $5$ TeV. This means that 
the composite Higgs boson is a tightly bound state, as if an elementary Higgs boson. 
In addition, relating to the top-quark Yukawa coupling $\bar g_{t}(\mu)$, 
the Yukawa couplings $\bar g_{b}(\mu)$ and $\bar g_{\tau}(\mu)$ also slowly vary 
from $1$ GeV to $5$ TeV, see Fig.~\ref{figyl}. These features 
imply that it should be hard to have any
detectable nonresonant signatures in the LHC $pp$-collisions, showing the deviations 
from the SM with the elementary Higgs boson. 
 
To end this article, we would like to mention that 
the vectorlike feature of $W^\pm$-boson coupling at high energy $\E$
is expected to have some collider signatures 
on the decay channels of $W^\pm$-boson into both left- and right-handed 
helicity states of two high-energy leptons or quarks. The branching ratios of
different helicity states are expected to be almost the same, given the qualitative 
estimation of Eqs.~(\ref{alphaw}) and (\ref{alphav}). 
This contrasts to the helicity suppression  
in the low-energy SM due to its $W^\pm$-boson coupling being purely 
left handed, recalling 
the helicity suppression of pion decay into an electron and the corresponding 
electron antineutrino. On the other hand,  
the forward-backward asymmetry in top-quark pair production measured 
by the CDF \cite{asCDF} and D0 \cite{asD0} at the Fermilab Tevatron $p\bar p$ collisions
seems to be larger than the SM (QCD) result. This may be related to 
the vectorlike (parity-restoration) feature of $W^\pm$-boson coupling at high energy, 
since the 
top-quark pair can be produced by $d,s$, and $b$ quarks in the $t$-channel via the $W^\pm$-boson
exchange. We will study it in detail.

\vskip0.1cm
\noindent
\section{\bf Acknowledgments}  
\hskip0.1cm The author thanks the anonymous referee for his/her effort of reviewing this article. The author 
thanks Professor Hagen Kleinert for discussions on the domains of IR- and UV-fixed 
points of gauge field theories, and to 
Professor Zhiqing Zhang for discussions on 
the experimental physics in the LHC.

\end{document}